\documentclass[10pt,aps,nobalancelastpage,prb,amsmath,twocolumn,superscriptaddress,showpacs]{revtex4-1}

\usepackage{graphicx}
\usepackage{amsmath}

\newcommand{\dvec}[1]{\ensuremath{\boldsymbol{#1}}}
\newcommand{\vk}{\dvec{\mathrm{k}}}
\renewcommand{\vr}{\dvec{\mathrm{r}}}

\begin{document}
\title{Electron Correlations in Bilayer Graphene}
\author{D. S. L. Abergel} 
\affiliation{Department of Physics and Astronomy, University of
Manitoba, Winnipeg, Canada R3T 2N2}
\affiliation{Condensed Matter Theory Center, University of Maryland,
College Park MD, 20742-4111}
\author{Tapash Chakraborty}
\email{tapash@physics.umanitoba.ca}
\affiliation{Department of Physics and Astronomy, University of
Manitoba, Winnipeg, Canada R3T 2N2}

\begin{abstract}
The nature of electron correlations in bilayer graphene has been
investigated. An analytic expression for the radial distribution function
is derived for an ideal electron gas and the corresponding
static structure factor is evaluated. We also estimate the interaction 
energy of this system.
In particular, the functional form of the pair-correlation function was
found to be almost insensitive to the electron density in the
experimentally accessible range. 
The inter-layer bias potential also has a negligible effect on the
pair-correlation function.
Our results offer valuable insights into the general behavior of the
correlated systems and serve as an essential starting-point for
investigation of the fully-interacting system.
\end{abstract}

\maketitle

Despite intense studies for many decades, the important role of
many-particle correlations in electron liquids \cite{mahan},
particularly in systems with reduced dimensions, remains a
challenging issue in condensed-matter physics.  This subject has become
even more pressing in recent years as the physical properties of
graphene have been unmasked at a rapid pace \cite{review}. Monolayer and
bilayer graphene are totally new classes of two-dimensional electron
systems with unusual band structures and chiral charge carriers. The
influence of electron correlations on various physical properties of the
chiral two-dimensional electron gas in monolayer graphene has been the
subject of several investigations \cite{polini,mishchenko,vafek}. 
These were all carried out with the aid of established pertubative
techniques.
On the other hand, our earlier work based on exact analytical treatment
\cite{abergel} indicated that the electron correlations completely
vanish from the two-particle kinetic energy of monolayer graphene, a
fact which was attributed to the specific spinor structure of the
single-particle wave functions which in turn is a direct manifestation
of the chirality of the massless Dirac fermions in monolayer graphene.
Such cancellations were not found to occur for electrons in bilayer
graphene \cite{abergel}, due to the massive chiral nature of the
low-energy quasiparticles. 
Clearly, we have a long way to go in order to find a satisfactory
understanding of the role interactions play in these unique electron
systems, but it is evident that the effects of correlations in bilayer
graphene is an important and relevant issue.

In this Rapid Communication, we will lay out the foundation for the
process of establishing the behavior of the correlation function of
interacting electrons in bilayer graphene, which is an essential step
for evaluation of the thermodynamic properties of this system. We derive
an analytic expression for the pair-correlation function (PCF) of an
ideal electron system, and use it to compute the corresponding static
structure factor as a function of the electron density. We make a
detailed comparison of the PCF with the same quantity in a traditional
two-dimensional electron gas (2DEG), and compute the exchange energy for
the bilayer graphene system.
Evaluation of the PCF with full electron correlations included is
certainly a very arduous task \cite{chandre}, and has not yet been
attempted for bilayer graphene. Our present approach is an important and
necessary first step in describing a fully correlated system, but it
already provides valuable insights into the general behavior of these
functions.

Bilayer graphene has a hexagonal Brillouin zone where the low-energy
features are located near two inequivalent corners, called the K
points \cite{mccann-prl96,mccann-prb74}. There are four low-energy $\pi$ 
bands for each spin species in the vicinity of each K point.  In unbiased 
bilayer graphene the second and third of these bands (called the 
`low-energy branches') touch exactly at the charge-neutrality point, making 
intrinsic bilayer graphene a zero-gap semiconductor. There are two more 
bands (the `split' branches), each seperated from the low-energy branch 
by the inter-layer coupling parameter $\gamma^{}_1 \approx 0.4\,
\mathrm{eV}$.  
We label these bands as follows:
The conduction and valence bands are given by $\nu=c,v$; the branches
are $b=l,s$; the valleys are $\xi=K,K'$; and the spins are
$\sigma=\uparrow,\downarrow$. Adding the electron wave vector $\vk$, we
have the complete set of quantum numbers $\lambda = \{\nu^{}_\lambda,
b^{}_\lambda, \vk^{}_\lambda, \xi^{}_\lambda, \sigma^{}_\lambda\}$.
At half-filling, all eight valence bands are filled, and
the eight conduction bands are empty. 

%The radial distribution function of an $N$-particle system is defined as 
%\cite{mahan}
%\begin{equation}
	%g(\vr^{}_1,\vr^{}_2) = \mathcal{A}^2 \int d^2\vr^{}_3\ldots d^2\vr^{}_N
	%\left| \Psi_{\lambda^{}_1,\ldots,\lambda^{}_N} 
		%\left( \vr^{}_1,\ldots,\vr^{}_N \right) \right|^2
%\end{equation}
%where $\Psi$ is the normalized $N$-fermion wave function which can be 
%expressed for the ideal case as a Slater determinant of single-electron
%states $\phi^{}_{\lambda}(\vr)$.
Spin- and valley-dependent contributions to the PCF are defined as
\cite{mahan}
\begin{equation}
	g^{}_{\sigma_\lambda\xi_\lambda,\sigma_\mu\xi_\mu} (\vr_1,\vr_2)
	= 1 - \frac{\left| \left\langle
	\Psi^{\dagger}_{\sigma_\lambda\xi_\lambda}(\vr_1) 
	\Psi^{}_{\sigma_\mu\xi_\mu}(\vr_2) \right\rangle \right|^2} 
	{n^{}_{\sigma_\lambda\xi_\lambda}(\vr_1) n^{}_{\sigma_\mu\xi_\mu}(\vr_2)}
	\label{eq:gFinitial}
\end{equation}
where $\Psi^{}_{\sigma_\lambda\xi_\lambda}(\vr) = 
\sum_{\vk_\lambda,\nu_\lambda,b_\lambda} \phi^{}_\lambda a^{}_\lambda$ is the 
field operator for an electron in valley $\xi_\lambda$ with spin
$\sigma_\lambda$. 
The total PCF can be expressed in terms of these
functions as $g(\vr_1,\vr_2) =
\sum_{\sigma_\lambda,\sigma_\mu,\xi_\lambda,\xi_\mu} \frac{1}{16}
g^{}_{\sigma_\lambda\xi_\lambda,\sigma_\mu\xi_\mu}(\vr_1,\vr_2)$.

To evaluate this expression we substitute the well-known form of the
single-particle wave functions in bilayer graphene
\begin{equation}
	\phi^{}_\lambda = \frac{e^{i\vk^{}_\lambda\cdot\vr}}{\sqrt{\mathcal{A}}}
	\Xi^{}_\lambda \otimes \Sigma^{}_\lambda \otimes
	\begin{pmatrix} e^{-i\theta^{}_\lambda} w^{}_\lambda \\ 
	e^{i\theta^{}_\lambda} x^{}_\lambda \\ y^{}_\lambda \\ z^{}_\lambda
	\end{pmatrix}
	\equiv \frac{e^{i\vk^{}_\lambda\cdot\vr}}{\sqrt{\mathcal{A}}}
	\Xi^{}_\lambda \otimes \Sigma^{}_\lambda \otimes \Phi^{}_\lambda
	\label{eq:phidef}
\end{equation}
where $\Xi^{}_\lambda$ and $\Sigma^{}_\lambda$ are respectively the
valley and spin parts of the wave function, $\mathcal{A}$ is the
normalization area, and the functional form of the wave function
components and single-particle energies are easily derived from the
Schr\"odinger equation for the tight-binding Hamiltonian.
Substituting Eq.~\eqref{eq:phidef} into Eq.~\eqref{eq:gFinitial}, we
find that
\begin{multline}
	g^{}_{\sigma_\lambda\xi_\lambda,\sigma_\mu\xi_\mu}(\vr) = 1 - 
	\frac{16 \delta_{\sigma_\lambda,\sigma_\mu}
	\delta_{\xi_\lambda,\xi_\mu}}
	{\mathcal{A}^2 n^2} \times \\
	\times \sum_{\substack{\vk_\lambda \nu_\lambda b_\lambda\\ \vk_\mu \nu_\mu
	b_\mu}}
	f^{}_\lambda f^{}_\mu \cos\left[ (\vk^{}_\lambda -
	\vk^{}_\mu) \cdot \vr \right] \Phi^\ast_\lambda \Phi^{}_\mu \cdot
	\Phi^\ast_\mu \Phi^{}_\lambda
	\label{eq:gsxsubs}
\end{multline}	
with $\vr = \vr^{}_1 - \vr^{}_2$ and where $f^{}_\lambda$ is the
occupancy of state $\lambda$. 
As expected, the off-diagonal components of the PCF
are constant with unit value. We have also assumed that
$n^{}_{\sigma_\lambda \xi_\lambda}(\vr_1) = n/4$ (i.e. that
electrons are equally distributed between the valley and spin
components and that electron density is uniform in space).

To procede, we must be careful about how we define the various
densities. The total density of electrons is
denoted by $n$, but we also consider the density of charge carriers
(also called the excess density)
$n^{}_\mathrm{cc} = n - n^{}_0$ which may be either positive (for
electrons) or negative (for holes). Then, the sums over occupied wave
vector states must be taken independently for each combination of band
and branch quantum numbers. Taking the limit of an infinite system (with
the electron density held constant) means that we can replace the sums
over wave vectors with two-dimensional integrals.
The integrals which result from this procedure are not automatically
convergent for large wave vectors. Therefore, we must introduce a
cut-off wave vector using some physical reasoning.
Consideration of the lattice structure shows that each unit cell
contributes four $\pi$ electrons (one per carbon atom), so that the
density of electrons at half-filling is $n^{}_0 = 8/(\sqrt{3}a^2)$ where
$a\approx
2.46\text{\AA}$  is the lattice constant. Therefore, we can set the wave
vector cut-off $\Lambda$ because 
\begin{equation*}
	\frac{\mathcal{A}}{4\pi^2} \int d^2\vk = \frac{N_0}{8}
	\,\Rightarrow\, 
	\frac{\Lambda^2}{4\pi} = \frac{n^{}_0}8 
	\,\Rightarrow\quad
	\Lambda = \frac{ 2\sqrt{ \pi }}{\sqrt[4]{3}a}
	\label{eq:Lambdadef}
\end{equation*}
where $N_0$ is the total number of electrons at half-filling. 
\begin{widetext} 
As an example, in intrinsic graphene (where the Fermi energy is exactly
at the charge neutrality point) the valence bands are all filled and the
conduction bands are all empty.  Therefore, the sum over bands and
branches in Eq.~\eqref{eq:gsxsubs} becomes 
\begin{equation*}
	g^{}_{\sigma_\lambda\xi_\lambda,\sigma_\mu\xi_\mu}(\vr)
	= 1 - \frac{ \delta_{\sigma^{}_\lambda,\sigma^{}_\mu}
			\delta_{\xi^{}_\lambda,\xi^{}_\mu} }
		{\pi^4 n^2 }
		\sum_{b^{}_\lambda, b^{}_\mu}
		\int_0^{2\pi} d\theta^{}_\lambda 
		\int_0^{2\pi} d\theta^{}_\mu
		\int_0^\Lambda k^{}_\lambda \, dk^{}_\lambda
		\int_0^\Lambda k^{}_\mu \, dk^{}_\mu
	\cos\left[ (\vk^{}_\lambda - \vk^{}_\mu) \cdot \vr \right]
		\Phi_\lambda^\ast \Phi^{}_\mu \cdot \Phi_\mu^\ast
		\Phi^{}_\lambda.
\end{equation*}
Using the expressions for the wave functions in Eq.~\eqref{eq:phidef}
%Eqs.~\eqref{eq:w}--\eqref{eq:epsilon} 
and evaluating the elementary integrations over the angles, we arrive at 
\begin{multline}
	g_{\sigma^{}_\lambda \xi^{}_\lambda \sigma^{}_\mu, \xi^{}_\mu}(\vr) =
	1 - \frac{4 \gamma_1^2 \delta_{\sigma^{}_\lambda,\sigma^{}_\mu}
			\delta_{\xi^{}_\lambda,\xi^{}_\mu} }
		{\pi^2 n^2 \hbar^2 v_F^2 }
	\Bigg\{
		\left( \int_0^\Lambda k J_0(kr) \left[w^2 \big|_{b=s} +
			w^2 \big|_{b=l}\right] dk \right)^2 
		+ \left( \int_0^\Lambda k J_0(kr) \left[ x^2 \big|_{b=s} +
			x^2 \big|_{b=l} \right] dk \right)^2  \\
		+ \left( \int_0^\Lambda k J_0(kr) \left[ y^2 \big|_{b=s} +
			y^2 \big|_{b=l} \right] dk \right)^2 
		+ \left( \int_0^\Lambda k J_0(kr) \left[ z^2 \big|_{b=s} +
			z^2 \big|_{b=l} \right] dk \right)^2 \\
		+ 2 \left( \int_0^\Lambda k J_0(kr) \left[ yz \big|_{b=s} +
			yz \big|_{b=l} \right] dk \right)^2
		+ 2 \left( \int_0^\Lambda k \left[ J_0(kr) - \frac{2}{kr}J_1(kr)
			\right] \left[ wx \big|_{b=s} + wx \big|_{b=l} \right] dk
			\right)^2 \\
		+ 2 \left( \int_0^\Lambda k J_1(kr) \left[ wy \big|_{b=s} +
			wy \big|_{b=l} \right] dk \right)^2
		+ 2 \left( \int_0^\Lambda k J_1(kr) \left[ wz \big|_{b=s} +
			wz \big|_{b=l} \right] dk \right)^2 \\
		+ 2 \left( \int_0^\Lambda k J_1(kr) \left[ xy \big|_{b=s} +
			xy \big|_{b=l} \right] dk \right)^2
		+ 2 \left( \int_0^\Lambda k J_1(kr) \left[ xz \big|_{b=s} +
			xz \big|_{b=l} \right] dk \right)^2 \Bigg\}
			\label{eq:Tintrinsic}
\end{multline}
\end{widetext}
where all terms are evaluated with $\nu=v$, 
and $J_0(x)$ and $J_1(x)$ are the zeroeth order and first order
cylindrical Bessel function respectively.
In the case of positively-doped graphene (where the charge carriers are
holes and the Fermi energy is in
the valence band), we assume that for moderate densities only the
low-energy band is depopulated \cite{splitocc} so that the lower 
integration limit becomes the Fermi wave vector $k^{}_F = \sqrt{\pi
n^{}_\mathrm{cc}}$ when $b=l$. 
For negatively-doped graphene, each squared term in
Eq.~\eqref{eq:Tintrinsic} gains a contribution from the low-energy
conduction band ($\nu=c$, $b=l$) with the Fermi wave vector replacing
$\Lambda$ as the upper limit in this integral.

\begin{figure}[tbp]
	\centering
	\includegraphics[]{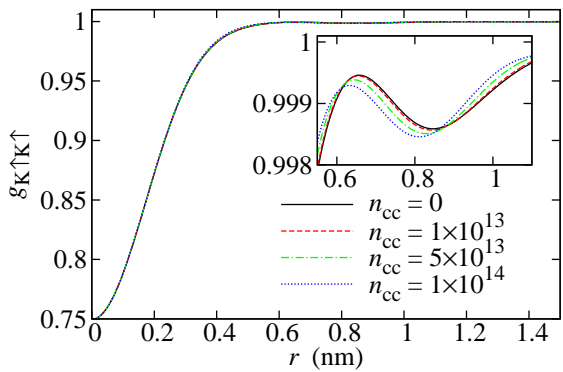}
	\caption{a) The pair-correlation function for several values of the
	electron density density. Black solid line: $n_\mathrm{cc}=0$; red
	dashed line: $n_\mathrm{cc}=10^{13}\mathrm{cm}^{-2}$; 
	green dash-dot line: $n_\mathrm{cc}=5\times10^{13}\mathrm{cm}^{-2}$;
	blue dotted line: $n_\mathrm{cc}=10^{14}\mathrm{cm}^{-2}$.
	b) Pair-correlation function near the exclusion hole edge. Densities
	are as in a).
	c) Comparison with 2DEG PCF. Red dashed line: 2DEG
	$n=10^{13}\mathrm{cm}^{-2}$; black solid line: 2DEG
	$n=7.6\times10^{15}\mathrm{cm}^{-2}$; black dotted line: bilayer
	$n_\mathrm{cc}=0$.
	\label{fig:gF}}
\end{figure}

In order to obtain the PCF, these integrals are
evaluated numerically and the resulting function is plotted in
Fig.~\ref{fig:gF} for various densities. 
The behavior of the function is clearly similar to that in a
conventional 2DEG \cite{glasser}, with an exchange hole with
radius approximately $5\text{\AA}$. 
The reason that $g(0)$ is finite is explained as follows: The PCF
evaluated here specifies all-but-two quantum numbers. Therefore for any
given combination of band and branch, there can be an electron at $r=0$
with one of three other combinations which does not violate the Pauli
principle. Hence the minimum value of the PCF is $3/4$, as seen in
Fig~\ref{fig:gF}. 
If we were to calculate the $g(r)$ for a fully-specified combination of
valley, spin, band and branch then this function would indeed go to zero
at the origin, just as it does for the conventional 2DEG.

The dependence of the PCF on the density is tiny for physically
reasonable values of the excess density. 
The reason for this tiny variation is that the electrons in the filled
valence bands contribute more to the sum over states than those in the
partially filled conduction band. The PCF
contains essentially an average over all particles $\frac{1}{N}\sum_i$
(where $N$ is the total number of electrons and $i$ runs over all filled
states).
The intrinsic density of electrons due to the valence bands which are
filled in the charge-neutral case is 
$n_0\approx 7.6\times 10^{15}\mathrm{cm}^{-2}$, which is much
greater than the density of charge carriers $n_\mathrm{cc}\lesssim
10^{14}\mathrm{cm}^{-2}$ due to the excess density induced by gating or
doping. Therefore when the average over all states is taken, the effect
of the partially filled conduction band (or partially empty valence
band) is swamped by the contribution from the filled valence band. 
This effect is highlighted by comparison with the non-interacting PCF in
a traditional semiconductor 2DEG (in the lower inset to
Fig.~\ref{fig:gF}). When the 2DEG PCF is plotted for
$n=1\times10^{13}\mathrm{cm}^{-2}$, the exchange hole is much larger
than in graphene. But when the total density
$7.6\times10^{15}\mathrm{cm}^{-2}$ is used the exchange hole is of a
much more similar size.
%This marks an important difference from the conventional 
%2DEG where the absence of contributions from the filled valence band 
%means that the effects of the excess electron density are the only 
%contributions to the PCF (and in fact, the 
%PCF is not defined for zero density in this case).

\begin{figure}[tbp]
	\centering
	\includegraphics[]{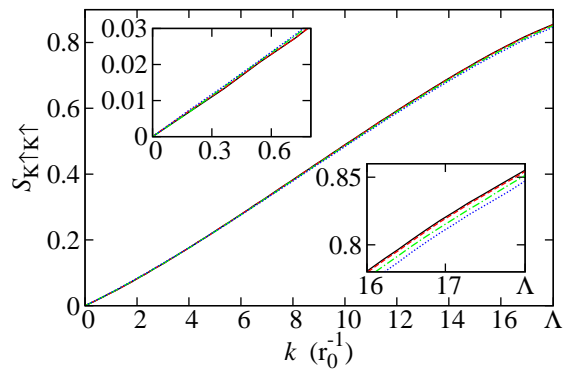}
	\caption{The static structure function for the same densities as in
	Fig.~\ref{fig:gF}a). The units are $r^{}_0 = \hbar
	v^{}_F/\gamma^{}_1 \approx 1.65 \mathrm{nm}$.
	Main plot: The whole wave vector range. 
	Upper-left inset: The low-wave vector region. 
	Lower-right inset: The wave vector region near the cut-off.
	\label{fig:Sk}}
\end{figure}

Once we obtain the radial distribution function, the static structure
factor for the system can be derived from the following expression
\cite{mahan}
\begin{equation*}
	S^{}_{\sigma_\mu\xi_\mu,\sigma_\lambda\xi_\lambda}(k) 
	= 1 + n^{}_{\sigma\xi}\int d^2\vr \left[
	g^{}_{\sigma_\mu\xi_\mu,\sigma_\lambda\xi_\lambda}(r)-1\right] 
	e^{-i\vk\cdot\vr} 
\end{equation*}
which is, in principle, an experimentally observable function via X-ray
and neutron diffraction where the correlation functions are usually
extracted from the measured diffraction intensity profile
\cite{roche-07}. We have evaluated the integral numerically and the
resulting function is plotted in Fig.~\ref{fig:Sk} for several values of
the electron density. We see that the variation with density is rather
small, but at low wave vector, $S(k)$ increases with density (upper-left
inset) while at high wave vector the opposite is true (lower-right
inset).  The structure factor is almost linear even up to the wave
vector cut-off $\Lambda$. This behavior has been noticed before in the
context of monolayer graphene \cite{polini,chandre}. 
This is noticably different from the result for the conventional 2DEG
\cite{glasser}, where the static structure function is roughly linear at
small wave vector but saturates at $S^{}_{\mathrm{2DEG}}(k) = \tfrac12$
for $k>2k^{}_F$. We emphasize that the static structure
function of bilayer graphene behaves similarly to the conventional 2DEG
at small wave vector, but like monolayer graphene at large wave vector.
This behavior might be expected when the quadratic-to-linear crossover
in the hyperbolic band structure is considered.

Finally, we calculate the exchange energy per electron associated
with the exchange-correlation hole
\begin{equation*}
	E^{}_{\rm int}(n) = \frac n2 \int d^2\vr V(r) \left[g(r)-1\right]
\end{equation*}
where $V(r)$ is the Coulomb potential and we use the full $g(r)$. 
This function is linear in the quasi-particle density, with $E(0)
\approx -2.5$\, eV.

Let us now turn our attention to the effect of a finite inter-layer bias
potential on the radial distribution function. When an electrostatic
potential is applied perpendicularly to the plane of the graphene, a gap
opens at the charge-neutrality point, and the shape of the low-energy
bands changes to a `Mexican hat' form 
\cite{mccann-prb74}.  This also changes the form of the wave functions,
and causes the Fermi surface
to become ring-shaped for small charge carrier density
\cite{stauber-prb75}.  Therefore, the integration limits in
Eq.~(\ref{eq:Tintrinsic}) change if the Fermi energy $E^{}_F < U/2$. In
that case, integrals relating to partially-filled bands become
$\int_0^{k^{}_F} dk \to \int_{k^{}_-}^{k^{}_+} dk$ with
\begin{equation*}
	k^{}_\pm = \sqrt{ \frac{4\pi^2 \hbar^2 v_F^2
	n_{\mathrm{cc}}^2}{U^2+\gamma_1^2} + 
	\frac{U^2}{4\hbar^2 v_F^2}
		\left(1 + \frac{\gamma_1^2}{U^2+\gamma_1^2}\right)
	\pm 2\pi n^{}_\mathrm{cc} }.
\end{equation*}
On the other hand, if $E^{}_F \geq U/2$ [which occurs when $k^{}_F >
U/(\hbar v^{}_F)$] then $k^{}_F = \sqrt{\pi
n^{}_\mathrm{cc}}$ as before.

\begin{figure}[tbp]
	\centering
	\includegraphics[]{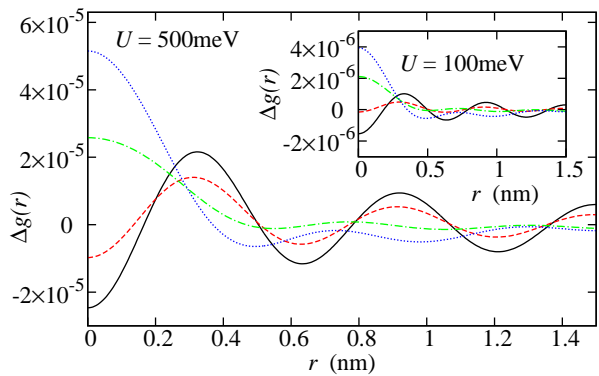}
	\caption{The change in $g(r)$ with the inter-layer bias potential
	for several densities. Main plot: $U=500\, \mathrm{meV}$; inset:
	$U=100\, \mathrm{meV}$. All densities are in $\mathrm{cm}^{-2}$ and
	line styles are the same as in Fig.~\ref{fig:gF}a).
	\label{fig:gfu}}
\end{figure}

We plot the change in the PCF with the
introduction of a bias $\Delta g = g^{}_U(r) - g^{}_0(r)$ as a function
of the inter-particle separation in Fig.~\ref{fig:gfu}.  We see that the
change is greatest at small charge carrier density and large $U$.
However, overall the change is very small which is predictable since the
PCF is related to the electron wave functions, and the inter-layer
potential only induces a change for $E\sim U \ll E(\Lambda)$. 
Similarly, the static structure factor shows only very small deviation
from the $U=0$ results for finite $U$.

In conclusion, we have investigated the PCF and the
corresponding static structure function for an ideal gas of electrons in 
bilayer graphene and compared it to the same quantity in the traditional
2DEG system. We have found behavior quite similar to that of the 
conventional 2DEG at equivalent density, in that an exchange hole is
formed with density-dependent radius. However, the manifestation of
effects due to the bands, especially the existence of the filled valence
band means that the dependence of these functions on
the density of charge carriers is minimal in the experimentally
accessible range. 
We have evaluated these functions for the gapped system as well, and
found that the effect of the inter-layer bias potential on these
quantities was also negligable.
This general picture will also be true for all Dirac-like systems which
have filled valence bands.
In the case when the many-body correlations are taken into account, we
expect very similar behavior for the dependence on density 
because electron-electron interactions do not alter the situation of
filled valence bands.
In monolayer graphene, we previously expected that the functional form
of the PCF remain insensitive to electron density in order to explain
the observed behavior of the electron compressibility \cite{abergel}.
It is interesting to observe a similar situation in the case of another
Dirac-like graphene system based on very general considerations. 

This work was supported by the Canada Research Chairs Programme.


\begin{thebibliography}{}
\bibitem{splitocc}
The split band is unoccupied while $E_F < \gamma_1$. The wave vector at
which this band begins to be occupied (at zero temperature) is $k_s =
\sqrt{2}\gamma_1/(\hbar v_F)$ which corresponds to a density of
$2.3\times 10^{14} \mathrm{cm}^{-2}$.
\end{thebibliography}
\end{document}